# Structural and magnetic properties of (Mg,Co)₂W hexaferrites

S.H. Mahmood[1*], Q. Al-Shiab[1], I. Bsoul[2], Y. Maswadeh[3], A. Awadallah[1]

[1]Physics Department, The University of Jordan, Amman 11942, Jordan

[2]Physics Department, Al al-Bayt University, Mafraq 13040, Jordan

[3]Central Michigan University, Mount Pleasant 48859, Michigan, United States of America

## Abstract

Precursor powders of $BaMg_{2-x}Co_xFe_{16}O_{27}$ with ($x = 0.0$, 1.0, and 2.0) were prepared using high-energy ball milling, and the effects of chemical composition and sintering temperature on the structural and magnetic properties were investigated using x-ray diffractometer (XRD), scanning electron microscopy (SEM), and vibrating sample magnetometry (VSM). XRD patterns of the prepared samples indicated that crystallization of pure BaW hexaferrite phase was achieved at sintering temperature of 1300° C, while BaM and cubic spinel phase intermediate phases were obtained at lower sintering temperatures of 1100° C and 1200° C. SEM images revealed improvement of the crystallization of the structural phases, and growth of the particle size with increasing the sintering temperature. The magnetic data of the samples sintered at 1300° C revealed an increase of the saturation magnetization from 59.44 emu/g to 72.56 emu/g with increasing Co concentration ($x$) from 0.0 to 2.0. The coercive field $H_c$ decreased from 0.07 kOe at $x$ = 0.0, to 0.03 kOe at $x = 1.0$, and then increases to 0.09 kOe at $x = 2.0$. The thermomagnetic curves of the samples sintered at 1300° C confirmed the existence of the W-type phase, and revealed spin reorientation transitions above room temperature.

## 1. Introduction

Hexagonal ferrites (Also known as hexaferrites), were discovered in the 1950s at Philips Research Laboratories. Since then, the degree of interest in these ferrites has been increasing exponentially due to their cost effectiveness and suitability for a wide range of industrial and technological applications [1-8]. Different types of these ferrites, the most important of which are the M-, Y-, W-, Z-, X-, and U-type, were successfully syntesized, and found to exhibit large variations of the magnetic properties due to the differences in their magnetic structures and magnetocrystalline anisotropy [2, 9-14]. Structurally, all these ferrites are hexagonal with almost the same lattice parameter $a = 5.88$ Å, and significantly different lattice parameter $c$, depending on the sequence of stacking of the structural blocks (S, R, and T) in the unit cell [9, 15]. Specifically, the $c$-parameter for M-type barium hexaferrite with chemical formula of $BaFe_{12}O_{19}$ and unit cell composed of RSR*S* staking is typically 23.2 Å. On the other hand, the $c$-parameter for W-type hexaferrite with chemical formula $BaMe_2Fe_{16}O_{27}$, space group $P6_3/mmc$, and RSSR*S*S* stacking is 32.8 Å [9, 16].

The W-type structure suggests that its unit cell can be represented by a combination of M-type unit cell and two spinel (S) structural blocks (S = $Me_2Fe_4O_8$), where the saturation magnetization can be derived from the superposition of the saturation magnetization of M + 2S. Accordingly, the saturation magnetization of W-type is expected to be higher than that of M-type hexaferrite of 20 $\mu_B$ per formual unit (corresponding to ~ 100 emu/g) at 0 K [17]. In BaW ferrite, the additional S block has the composition $[Me_2Fe_4O_8]^0$, which is electrically neutral, and has a net magnetization depending on the Me cation. For example, the experimental value of $Fe_2W$ is 27.4 $\mu_B$ (corresponding to 98 emu/g at 0 K and 78 emu/g at 293 K), which is in good agreement with the theoretical value of 28 $\mu_B$. Also, the magnetic moment of $Mg_2W$ at 0 K is expected to be equal to that of BaM, since the magnetic moment of $Mg_2Fe_4O_8$ is theoretically zero, while the moment of $Cu_2W$ is expected to be higher (22 $\mu_B$). Deviations of the experimental values from the theoretical values are expected due to the possibility of $Me_2$ ions occupying sites of the R block, and measurements at fields lower than those required for full magnetic saturation [9].

The small metal cations (Fe and Me) in W-type hexaferrite reside ine seven different interstitial crystallographic sites known as $4f_{VI}$, $2d$, $12k$, $6g$, $4f$, $4f_{IV}$, and $4e$ [18, 19]. These crystallographic sites are normally grouped into five magnetic sublattices as shown in Table 1 [16, 20, 21].

**Table 1:** Crystallographic and magnetic sites and their coordinations, positions in the unit cell, net magnetic sublattice spin orientation, and occupancy for BaW structure.

| Magnetic site | Crystallographic Site | Coordination | Block | Spin | Number of metal ions |
|---|---|---|---|---|---|

| $f_{VI}$ | $4f_{VI}$ | Octahedral | R | Down | 2 |
|---|---|---|---|---|---|
| $a$ | $6g$ | Octahedral | S-S | Up | 3 |
| | $4f$ | Octahedral | S | Up | 2 |
| $f_{IV}$ | $4e$ | Tetrahedral | S | Down | 2 |
| | $4f_{IV}$ | Tetrahedral | S | Down | 2 |
| $k$ | $12k$ | Octahedral | R-S | Up | 6 |
| $b$ | $2d$ | Bi-pyramidal | R | Up | 1 |

Extensive research work on W-type hexaferrites prepared by different synthesis routes with a variety of cationic substitutional scenarios was carried out due to their potential for applications in microwave absorption [22-37], magnetic recording [20, 38-40], and other electrical devices [41-43]. All BaW hexaferrites, with the exception of $Co_2W$, are characterized by a uniaxial anisotropy. W-type hexaferrites containing $Co^{2+}$ ions exhibit a comples magnetic structure with the variation of temperature, where spin reorientation transitions from easy plane, to easy cone, to easy axis are expected, making these ferrites of potential importance for magnetic refrigration [24, 44-47] $Co_2W$ ferrite ($BaCo_2Fe_{16}O_{27}$) has a cone of easy magnetization with a constant vertex angle of 70° to the $c$-axis in the temperature range from -273° C to 180° C. As the temperature increases, the easy magnetization direction rotates towards the $c$-axis until the ferrite becomes uniaxial at 280°C [48]

Much of the research work on W-type hexaferrites available in the literature was concerned with Co- and Zn-based ferrites with a variety of cationic substitutions. Modification of the properties of the W-type ferrite based on $Mg^{2+}$ as the divalent metal ion, however, was not addressed sufficiently in the literature [49], especially, adequate structural and magnetic characterization. In this study, the effects of $Co^{2+}$ cationic substitution for $Mg^{2+}$ on the structural and magnetic properties of $BaMg_2W$ hexaferrites were investigated by XRD, SEM, VSM, and Mössbauer spectroscopy. The results of measurements made by the various experimental techniques were compared to reach an understanding of the crystalline and magnetic structure of the compounds. The spin reorientation transitions in the $Co^{2+}$-substituted compounds was investigated by means of thermomagnetic measurements

## 2. Experimental

Samples of BaMg$_{2-x}$Co$_x$Fe$_{16}$O$_{27}$ ($x$ = 0.0, 1.0 and 2.0) were prepared by ball milling stoichiometric ratios of high purity (~ 99%) barium carbonate (BaCO$_3$), Fe$_2$O$_3$, CoO, and MgO precursor powders using a high-energy ball mill (Fritsch Pulverisette-7) equipped with zirconia bowls and balls. The milling was carried out in an acetone medium for a period of 16 h at an angular speed of 250 rpm. The product was then left to dry in air at room temperature. A cylindrical pellets of about 1.2 cm in diameter and ~2 mm in thickness were made by pressing ~ 0.8 g of the powder under a 5-ton force in a stainless steel die. The discs were then sintered in an oven at 1100° C, 1200° C, and 1300° C for 2 h in air.

X-ray diffraction (XRD) patterns of the sintered samples were collected using XRD 7000-Shimadzu diffractometer with Cu-K$_\alpha$ radiation ($\lambda_1$ = 1.540560Å, $\lambda_2$ = 1.54439 Å), in the angular range 20° ≤ 2$\theta$ ≤ 70° with scanning step of 0.01. The patterns were analyzed using Expert High Score 2.0.1 software to identify the structural phases, and Rietveld analysis was performed for structural refinement using FullProf software.

SEM system (FEI-Inspect F50/FEG) equipped with energy dispersive spectrometer (EDS) was used to investigate the particle morphology and size distribution, as well as the homogeneity and local chemical composition of the prepared samples. The magnetic properties of the samples were examined using vibrating sample magnetometer (VSM Micro Mag 3900, Princeton Measurements Corporation), which operated at applied fields up to 10 kOe.

## 3. Results and discussion

### 3.1 XRD results

The XRD patterns were collected in the angular range 0° ≤ 2$\theta$ ≤ 70° for all samples BaMg$_{2-x}$Co$_x$Fe$_{16}$O$_{27}$ ($x$ = 0.0, 1.0, 2.0) sintered at 1100° C, 1200° C, and 1300° C. The XRD patterns of samples with different $x$ values, but sintered at the same temperature, were similar, and a representative set of XRD patterns for the sample with $x$ = 1 is shown in Fig. 1. XRD analysis using Expert High score software revealed that all samples sintered at temperatures below 1300°

C were multicomponent, consisting of BaM and cubic spinel MeFe$_2$O$_4$ (Me = Mg, Co) structural phases. The analysis also revealed that each samples sintered at 1300° C was single Mg$_2$W or Co$_2$W phase. These results indicated that crystallization of the W-type phase at 1300° C was preceded by the crystallization of the intermediate BaM and cubic spinel phases at lower temperatures.

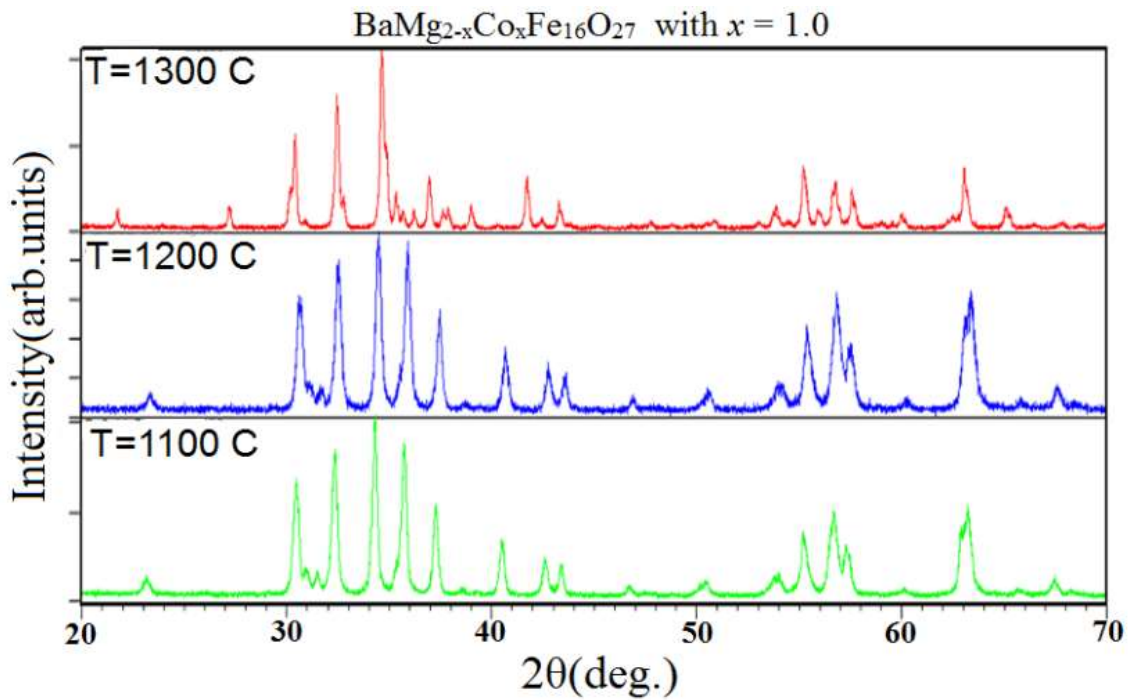

Fig. 1: X-ray diffraction patterns of BaMgCoFe$_{16}$O$_{27}$ ferrites sintered at different temperatures.

XRD pattern for the sample with $x = 2$ sintered at 1300° C revealed the existence of a single Co$_2$W (BaCo$_2$Fe$_{16}$O$_{27}$) hexaferrite phase consistent with the standard pattern (JCPDS 01-078-0135), while the patterns of the samples sintered at 1100° C and 1200° C indicated the coexistence of BaFe$_{12}$O$_{19}$ (BaM) hexaferrite phase consistent with the standard pattern (JCPDS 00-043-0002) together with CoFe$_2$O$_4$ spinel phase matching the standard pattern (JCPDS 01-079-1744). Also, the diffraction pattern for the sample with $x = 1.0$ sintered at 1300° C shows the existence of MgCo-W (BaMgCoFe$_{16}$O$_{27}$) hexaferrite phase consistent with the standard pattern (JCPDS 01-078-0135). The patterns for the sample sintered at 1100° C and 1200° C, however, indicated the existence of BaM hexaferrite phase consistent with the standard pattern (JCPDS 00-043-0002), together with $MgFe_2O_4$ spinel phase consistent with the standard pattern (JCPDS 01-089-3084), and $CoFe_2O_4$ spinel consistent with the standard pattern (JCPDS 01-079-1744). Similarly, the XRD pattern for the sample with $x = 0.0$ sintered at 1300° C indicated the existence of a single $Mg_2W$ ($BaMg_2Fe_{16}O_{27}$) hexaferrite phase consistent with the standard pattern (JCPDS 01-078-1551), while the patterns for the samples sintered at 1100° C and 1200° C revealed the existence of BaM hexaferrite phase consistent with the standard pattern (JCPDS 00-043-0002), together with Mg-spinel ($MgFe_2O_4$) phase consistent with the standard (JCPDS 01-089-3084). These results indicated that the M-phase and the spinel phase are intermediate phases that react to form the W-type phase at 1300° C.

Rietveld refinement of the XRD patterns of the samples sintered at 1300° C was performed using FullProf fitting routine (Fig. 2), and the refined lattice parameters and cell volume of the W phases are tabulated in Table 2. The lattice parameters $a$ and $c$ for $Co_2W$ are slightly lower than those for $Mg_2W$, in good agreement with previously reported results [18]. Also, the cell volume $V$ of $Co_2W$ is slightly ($< 0.1\%$) lower than that of $Mg_2W$. These results cannot be associated with the differences between the ionic radii of $Mg^{2+}$ and $Co^{2+}$ ions, since $Co^{2+}$ ions have a slightly higher ionic radius at octahedral site (0.745 Å) than $Mg^{2+}$ (0.72 Å), and the two ions have almost the same ionic radius at tetrahedral sites (0.57-0.58 Å) [50]. Accordingly, the variations of the structural parameters can be associated with lattice distortions [5].

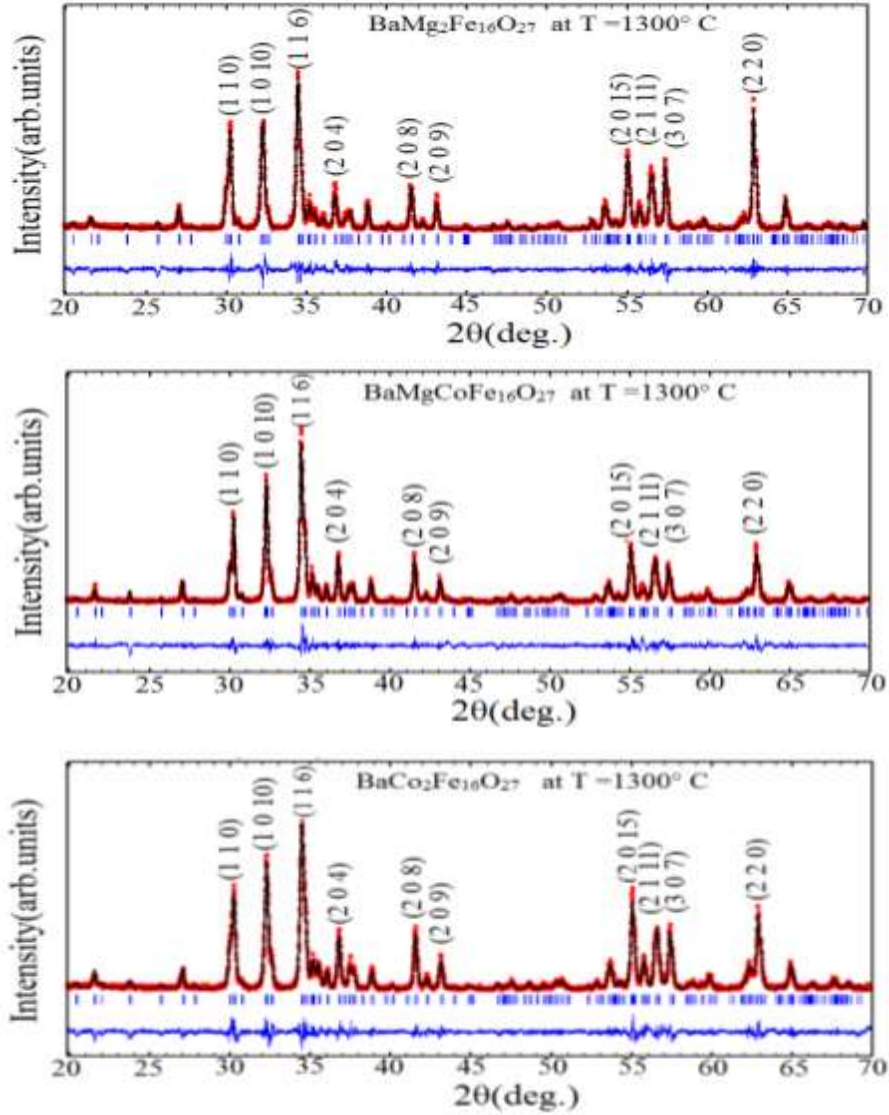

Fig. 2: Rietveld Refinement of the X-ray diffraction patterns of the system $BaMg_{2-x}Co_xFe_{16}O_{27}$ ($x$ = 0.0, 1.0, 2.0) sintered at 1300° C.

**Table 2:** Results of the refinement of the patterns for the samples $BaMg_{2-x}Co_xFe_{16}O_{27}$ ($x$ = 0.0, 1.0, 2.0) sintered at 1300° C.

| Phase formula | lattice parameters (Å) | | $V$ (Å$^3$) | $\rho_x$ (g/cm$^3$) | $R_F$ | $R_B$ | $\chi^2$ |
|---|---|---|---|---|---|---|---|
| | $a = b$(Å) | $c$( Å) | | | | | |
| $BaMg_2Fe_{16}O_{27}$ | 5.9076 | 32.9681 | 996.4 | 5.04 | 1.36 | 1.36 | 1.87 |
| $BaMgCoFe_{16}O_{27}$ | 5.9061 | 32.9565 | 995.6 | 5.15 | 0.67 | 0.89 | 1.97 |
| $BaCo_2Fe_{16}O_{27}$ | 5.9069 | 32.9552 | 995.8 | 5.27 | 0.73 | 1.04 | 1.91 |

The X-ray density ($\rho_x$) increased linearly with increasing $x$, as shown in Table 2, recording an increase of 4.56% for the sample with $x = 2.0$. This increase is mainly due to the increase of the molecular weight of the W-type hexaferrite by 4.58% when Co replaced Mg completely, since the effect of the decrease of the cell volume on the X-ray density is much smaller.

The bulk density ($\rho_b$) of the samples sintered at 1300° C were measured by Archimedes method, and the porosity ($P$) of each sample was evaluated using the relation:

$$P = 1 - \frac{\rho_b}{\rho_x} \qquad (3.3)$$

These porosity of all samples (Table 3) is relatively low, indicating the possibility of producing highly densified (> 90% dense) magnets using the present synthesis route.

**Table 3:** X-ray density ($\rho_x$), bulk density ($\rho_b$), and porosity ($P$) of BaMg$_{2-x}$Co$_x$Fe$_{16}$O$_{27}$ ($x = 0.0$, 1.0, and 2.0) sintered at 1300° C.

| $x$ | $\rho_b$ (g/cm$^3$) | $\rho_x$ (g/cm$^3$) | $P$ (%) |
|-----|---------------------|---------------------|---------|
| 0.0 | 4.72 | 5.04 | 6.4 |
| 1.0 | 4.75 | 5.15 | 7.7 |
| 2.0 | 4.82 | 5.27 | 8.5 |

Crystallite size ($D$), lattice strain, and instrumental effects are the main factors that cause the broadening of the diffraction peaks [51]. The instrumental broadening must be subtracted from the observed peak broadening in order to determine the effect of the lattice strains and crystallite sizes on the peak broadening. The instrumental broadening was estimated from the broadening of the diffraction peaks of a standard silicon sample. In our study, the effect of the lattice strain was found to be very small and it could be neglected. Therefore, the only effect of the broadening of the diffraction peaks was the crystallite-size.

The crystallite size was determined using the Stokes and Wilson equation [5]:

$$D = \frac{\lambda}{\beta \cos \theta} \qquad (3.4)$$

where $\theta$ is the peak position, $\beta$ is the integral breadth (= Area under the peak divided by the maximum intensity), and $\lambda$ is the wavelength of radiation (1.5406 Å). The integral breadth and peak position was determined by fitting a diffraction peak with a Lorentzian line shape, and the crystallite size along the corresponding crystallographic direction was evaluated. Analysis of the (1 1 0) peak at $2\theta = 30.4°$ was carried out to determine the crystallite size along the basal plane of the hexagonal lattice, while analysis of the (1 0 10) peak at $2\theta = 32.4°$ was performed to explore the crystallite size along the $c$-direction. Furthermore, the peak at $2\theta = 34.6°$ was considered to explore the crystallite size in the more general direction perpendicular to the corresponding (116) crystallographic planes. The values of $D$ for all samples along the different investigated directions revealed nanocrystalline nature for all samples with crystallite size between 32 nm and 40 nm (Table 4).

**Table 4:** The crystallite size for $BaMg_{2-x}Co_xFe_{16}O_{27}$ ($x = 0.0, 1.0, 2.0$) sintered at 1300° C evaluated along different crystallographic directions.

| | $D$ (nm) | | |
|---|---|---|---|
| $x$ | (1 1 0) | (1 0 10) | (1 1 6) |
| 0.00 | 36 | 34 | 35 |
| 1.00 | 58 | 57 | 40 |
| 2.00 | 30 | 44 | 32 |

## 3.2 SEM results

SEM image of $BaMg_{2-x}Co_xFe_{16}O_{27}$ with $x = 0.0$ sintered at 1100° C (shown in Fig. 3-a) indicated that the sample consisted mostly of cuboidal particles and hexagonal platelets with average size around 230 nm. The particles in this sample seem to agglomerate with relatively high inter-particle porosity. Also, SEM image of $BaMg_{2-x}Co_xFe_{16}O_{27}$ with $x = 0.0$ sintered at 1200° C (Fig.

3-b) indicated that the sample consisted of cuboidal particles and hexagonal platelets with a relatively wide distribution of particle size mostly in the range between 0.2-1.0 µm. The particles in this sample stacked with low inter-particle porosity.

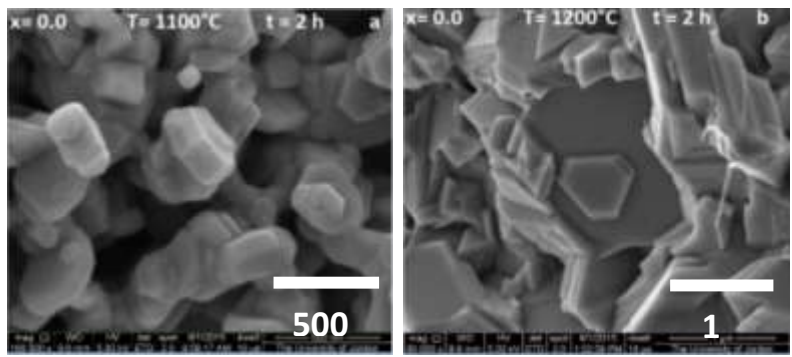

Fig. 3: SEM images of $BaMg_{2-x}Co_xFe_{16}O_{27}$ with $x = 0.0$ sintered at (a) 1100° C, and (b) 1200° C.

Representative SEM images of $BaMg_{2-x}Co_xFe_{16}O_{27}$ with $x = 0.0$ sintered at 1300° C (Fig. 4) revealed granular structure mainly composed of regular hexagonal plates characterized by a wider distribution of particle size. The diameters of the hexagonal plates ranged between 0.7-5 µm with average grain size of 2.2 µm. Since the critical single-domain size in hexaferrite was reported to be 0.46 micron [52], we may conclude that the sample consists of multi-domain particles. The particles in this sample stacked closely with relatively low porosity.

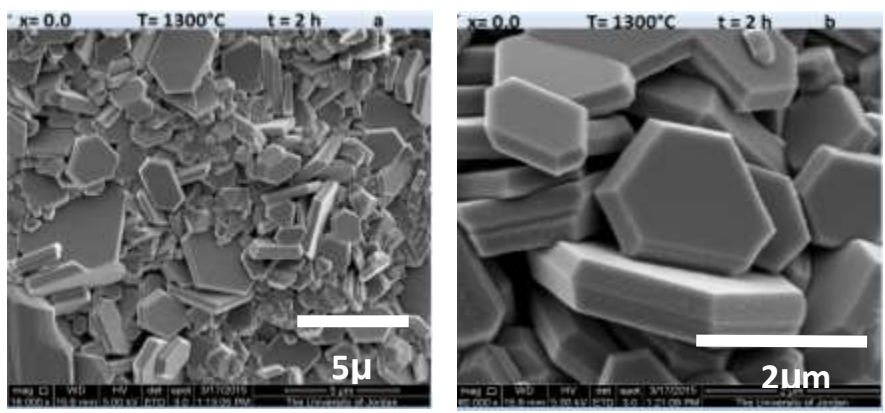

Fig. 4: SEM images for $BaMg_{2-x}Co_xFe_{16}O_{27}$ with $x = 0.0$ sintered at 1300° C.

SEM image of $BaMg_{2-x}Co_xFe_{16}O_{27}$ with $x = 1.0$ sintered at 1100° C (Fig. 5-a) also indicated that the sample consisted mainly of cuboidal particles and hexagonal platelets with diameters ranging between 100-500 nm. Improvement of the crystallization of hexagonal platelets, and increase of the particle size to the range of 0.5-1.7 μm was observed at sintering temperature of 1200° C (Fig. 5-b). Most of the particles in this sample are below the critical single domain size of about 0.5 μm, and a small fraction is > 1.0 μm in size.

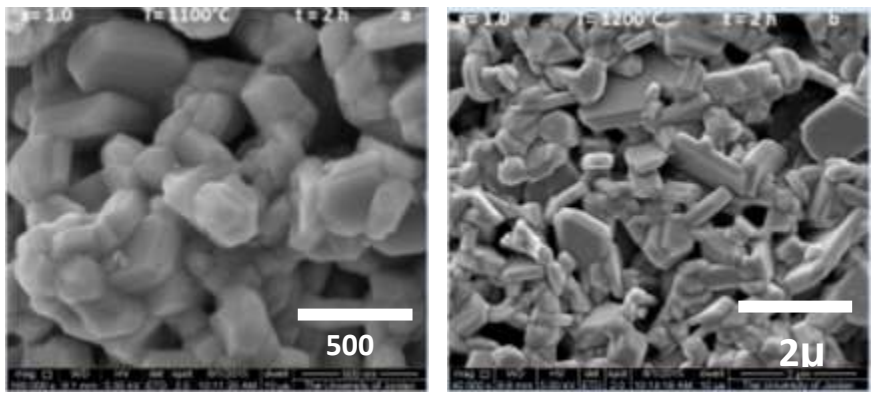

Fig. 5: SEM images of BaMg$_{2-x}$Co$_x$Fe$_{16}$O$_{27}$ with $x = 1.0$ sintered at (a) 1100° C, and (b) 1200° C.

Fig. 6 shows representative SEM images of BaMg$_{2-x}$Co$_x$Fe$_{16}$O$_{27}$ with $x = 1.0$ sintered at 1300° C. The sample is generally composed of non-granular mass, with only a small fraction of hexagonal plates. The diameters of the hexagonal plates were between 1-3 μm.

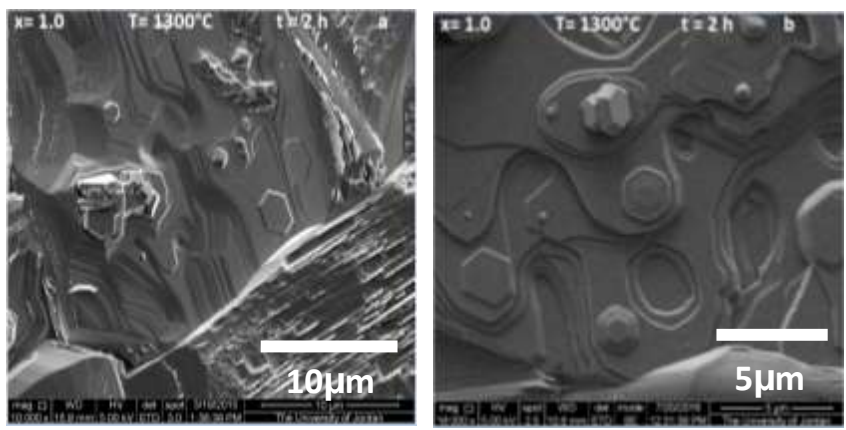

Fig. 6: SEM images of BaMg$_{2-x}$Co$_x$Fe$_{16}$O$_{27}$ with $x = 1.0$ sintered at 1300° C.

In addition, SEM images (Fig. 7) of $BaMg_{2-x}Co_xFe_{16}O_{27}$ with $x = 2.0$ sintered at 1100° C

indicated that the sample consisted mainly of cuboidal particles and hexagonal platelets with a relatively narrow distribution of particle size and average size around 240 nm. The sample sintered at 1200° C (Fig. 7-b) indicated improved crystallization of hexaginal particles, and the existence of large non-particulate masses.

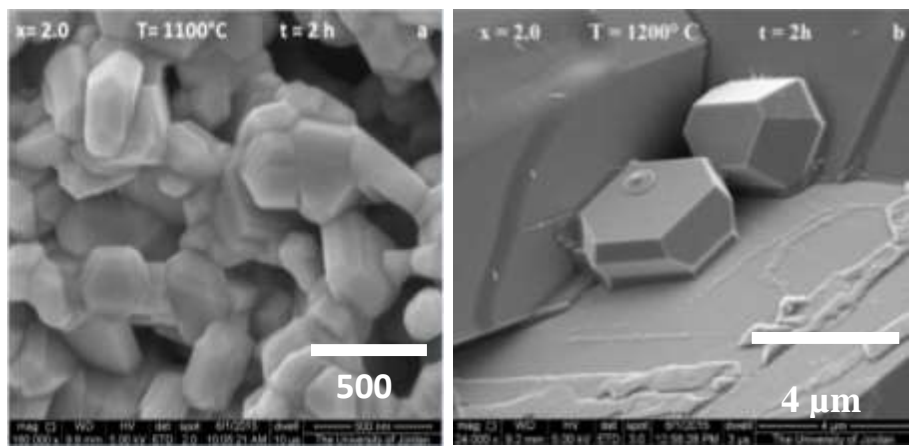

Fig. 7: SEM images of $BaMg_{2-x}Co_xFe_{16}O_{27}$ with $x = 2.0$ sintered at (a) 1100° C, and (b) 1200° C.

Fig. 8 shows SEM images of the $BaMg_{2-x}Co_xFe_{16}O_{27}$ with $x = 2.0$ sintered at 1300° C. The sample is generally composed of large non-granular masses, with only a small fraction of perfect hexagonal plates, the diameters of which were in the range of 0.5-2 μm. The non-granular masses developed in layered, nonporous structures.

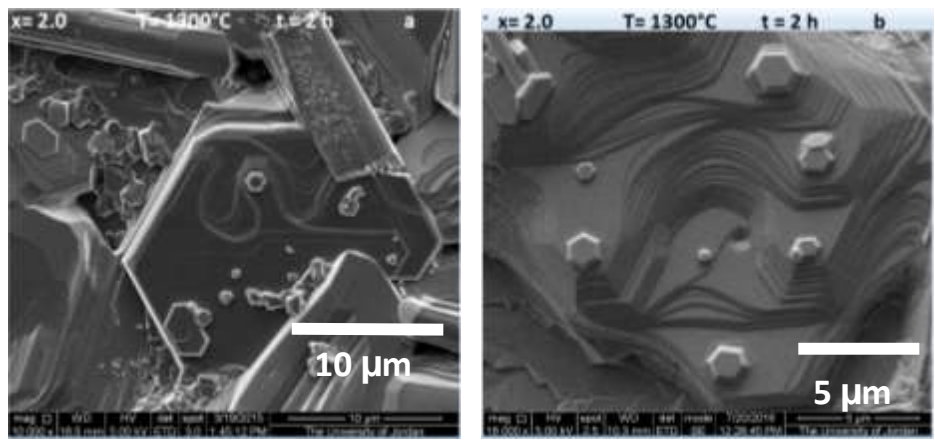

Fig. 8: SEM images of $BaMg_{2-x}Co_xFe_{16}O_{27}$ with $x = 2.0$ sintered at 1300° C.

Energy dispersive x-ray spectroscopy (EDS) was used to examine the chemical composition of the samples sintered at 1300° C. The spectra collected from two different spots of the sample with $x = 0.0$ revealed that the Fe:Ba ratio was 14.9 and 16.3, while the Mg:Ba ratio was 1.26 and 1.52. The Fe:Ba ratio is consistent with the theoretical stoichiometric ratios of 16.0 for $Mg_2W$ hexaferrite, while the difference between the observed Mg:Ba ratio and the theoretical ratio of 2 cannot be accounted for by the calculated experimental uncertainty (of ~ 20%). This discrepancy could be associated with the low signal for Mg, and the position of its emission line in a sloped and noisy region of the background, which makes the line intensity evaluated by the software unreliable.

 EDS spectra from two different spots of the sample with $x = 1.0$ indicated that the Fe:Ba ratio was 15.7 and 14.0, which is consistent with the theoretical value within the experimental uncertainty of ~ 17%. The Mg:Ba ratio at the two1.95 and 0.98.  Although the ratio from the second spot was in good agreement with the theoretical value of 1.00, the significantly higher value from the first spot may indicate unreliability of the evaluated Mg:Ba ratio due to the weak signal of Mg occurring in a rather noisy and sloped back ground. On the other hand, the Co:Ba ratio at the two spots was 2.58 and 2.06, which is more than double the theoretical value of 1.00. This difference cannot be accounted for by the experimental uncertainty of ~ 19%, and the observed high Co:Ba ratio can be attributed to the overlapping between the Co-$K_\alpha$ line (which was used for evaluation of the Co concentration) and the Fe-$K_\beta$ spectral line (the difference of the energies of these two lines is < 2%).

EDS spectrum collected from a representative point in the sample with $x = 2.0$ revealed that Fe:Ba = 15.6, which is in good agreement with the theoretical value of 16.0. On the other hand, Co:Ba ratio was found to be 3.86, which is significantly higher than the theoretical value of 2.00. Again, this high value is associated with the apparent increase of the spectral intensity of the Co-$K_\alpha$ line due to overlapping with Fe-$K_\beta$ spectral line.

## 3.3 VSM results
### 3.3.1 Room temperature hysteresis loop

In this section, the magnetic properties of the system $BaMg_{2-x}Co_xFe_{16}O_{27}$ ($x = 0.0$, 1.0, and 2.0) samples sintered at 1100° C, 1200° C, and 1300° C, were investigated using room temperature hysteresis loop (HL) measurements in an applied field up to 10 kOe. The hysteresis loops (HLs) for all the samples are shown in Fig. 9, 10, and 11. The values of the saturation magnetization, $M_s$, remanence magnetization, $M_r$, and coercive field, $H_c$, were determined from the hysteresis loops.

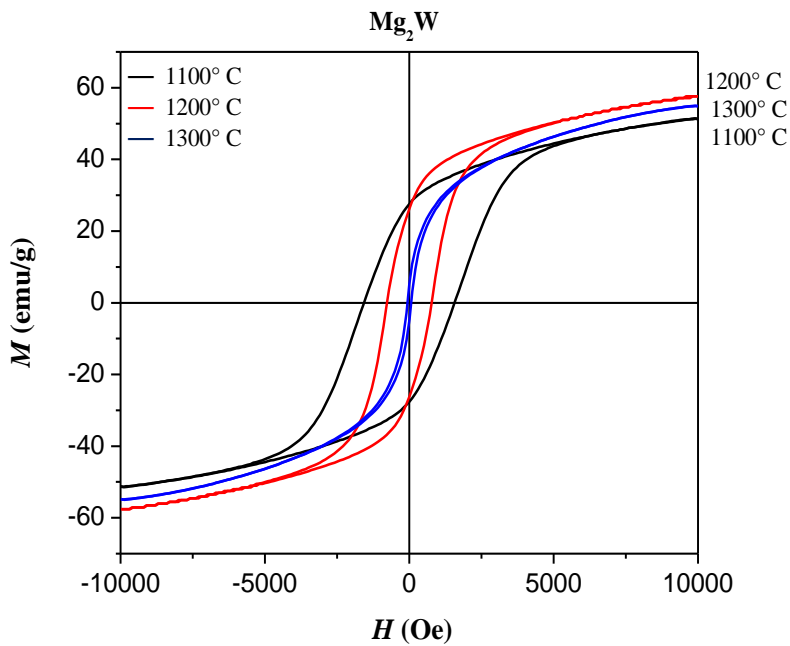

Fig. 9: Hysteresis loops of $BaMg_{2-x}Co_xFe_{16}O_{27}$ with $x = 0.0$ sintered at 1100° C, 1200° C, and1300° C.

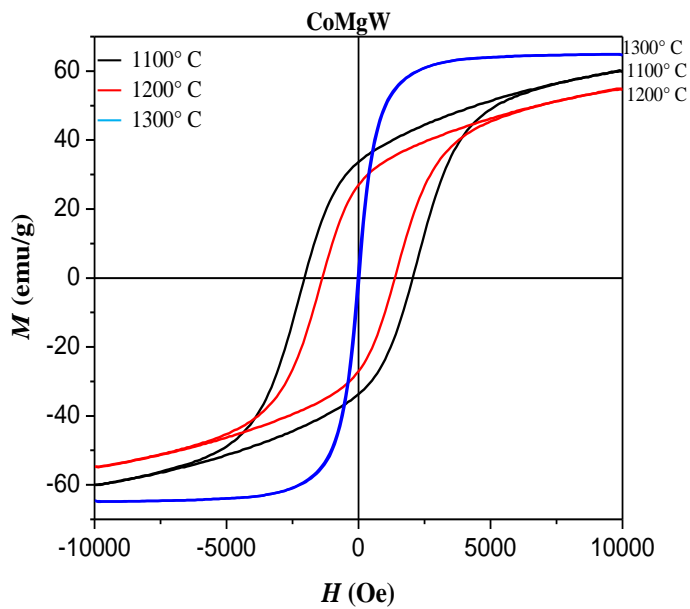

Fig. 10: Hysteresis loops of BaMg$_{2-x}$Co$_x$Fe$_{16}$O$_{27}$ with $x$ = 1.0 sintered at 1100° C, 1200° C, and1300° C.

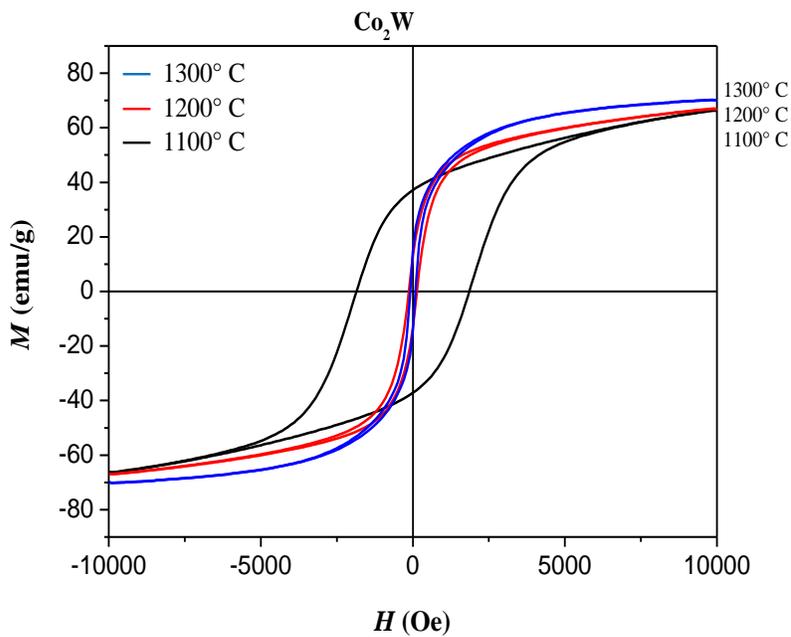

Fig. 11: Hysteresis loops of BaMg$_{2-x}$Co$_x$Fe$_{16}$O$_{27}$ with $x$ = 2.0 sintered at 1100° C, 1200° C, and1300° C.

The HLs for all samples indicated magnetic softening with the increase of the sintering temperature. The coercivity and remanence magnetization were determined directly from the hysteresis loops, while the saturation magnetization, $M_s$, was determined from the law of approach to saturation (LAS) [12, 53]:

$$M = M_s \left( 1 - \frac{A}{H} - \frac{B}{H^2} \right) + \chi H \tag{3.5}$$

Here $M$ is the magnetization (in emu/cm$^3$), $M_s$ is the spontaneous saturation magnetization of the domains per unit volume, $A$ is a constant associated with microstress and/or inclusions, $B$ is a constant representing the magnetocrystalline anisotropy contribution, and $\chi H$ is the forced magnetization term. Plotting $M$ versus $1/H^2$ in the high field range $8.5 - 10$ kOe gave straight lines, indicating that the magnetization in this field range is determined completely by the magnetocrystalline term.

The saturation magnetization was determined from the intercept of the straight line with the $M$-axis. The magnetic parameters derived from the hysteresis loops are tabulated in Table 5.

**Table 5:** Saturation magnetization, remanence magnetization, squareness ratio, and coercive field for BaMg$_{2-x}$Co$_x$Fe$_{16}$O$_{27}$ sintered at different temperatures.

| $x$ | $T(^\circ$ C$)$ | $M_s$(emu/g) | $M_r$(emu/g) | $M_{rs}$ | $H_c$(kOe) |
|---|---|---|---|---|---|
| 0.0 | 1100 | 55.40 | 27.54 | 0.50 | 1.57 |
| 0.0 | 1200 | 62.1 | 26.06 | 0.42 | 0.77 |
| 0.0 | 1300 | 59.44 | 4.33 | 0.07 | 0.07 |
| 1.0 | 1100 | 64.5 | 33.60 | 0.52 | 2.05 |
| 1.0 | 1200 | 59.84 | 26.91 | 0.45 | 1.37 |
| 1.0 | 1300 | 65.22 | 2.66 | 0.04 | 0.03 |
| 2.0 | 1100 | 72.60 | 37.11 | 0.51 | 1.85 |
| 2.0 | 1200 | 71.23 | 12.09 | 0.17 | 0.14 |
| 2.0 | 1300 | 72.56 | 11.67 | 0.16 | 0.09 |

Because the W-type hexaferrite structure is built up of M-type hexaferrite and S block, where S block contains two spinel molecules, the magnetic moment of the W-type hexaferrite can be described by the relations [9, 34, 54],

$$\mu_W = \mu_M + 2\,\mu_S \qquad (3.8)$$

$$\mu = \frac{M M_s}{N_A \mu_B} \qquad (3.9)$$

Where $\mu_M$ is the magnetic moment of $BaFe_{12}O_{19}$, and $\mu_S$ is the magnetic moment of spinel, $M$ is molecular weight, $M_s$ is the saturation magnetization, $N_A$ is Avogadro's number ($6.023 \times 10^{23}$ molecules/mole), $\mu_B$ is the Bohr magneton ($9.27 \times 10^{-21}$ erg/G), and $\mu$ is the magnetic moment of a given compound. Therefore by substituting equation (3.9) into (3.8) we obtain:

$$(M_s)_W = \left[\frac{M_M}{M_W}\right](M_s)_M + 2\left[\frac{M_S}{M_W}\right](M_s)_S \qquad (3.10)$$

This last equation predicts that the saturation magnetization of $Mg_2W$ is 58.62 emu/g, which is in good agreement with the observed value of 59.44 emu/g for $Mg_2W$ sample sintered at 1300° C. On the other hand, the expected theoretical saturation magnetization of MgCoW is 66 emu/g, which is also in good agreement with the observed value of 65.22 emu/g. Further, the expected saturation magnetization of $Co_2W$ is 72.96 emu/g, which is also in good agreement with the observed value of 72.56 emu/g. These results indicate that the magnetization of the W-type phase can be determined from the superposition of the magnetizations of the underlying magnetic phases.

The saturation magnetization for the sample with $x = 0$ increased with sintering temperature from 55.4 emu/g for the sample sintered at 1100° C, to 62.1 emu/g for the samples sintered at 1200° C. The sample sintered at 1100° C is a mixture of BaM phase and $MgFe_2O_4$ phase as indicated by XRD analysis. Assuming that the magnetization of the sample results from a simple superposition of the BaM and Mg-spinel phases, and using the values of the saturation magnetizations of polycrystalline BaM phase of 70 emu/g [1], and of 27 emu/g for $MgFe_2O_4$ phase [9], the saturation magnetization for the sample sintered at 1100° C (55.4 emu/g) suggests that this sample is composed of about 66 wt. % BaM phase and 34 wt. % $MgFe_2O_4$ phase. If all Mg were incorporated in the formation of $MgFe_2O_4$ spinel phase, the sample would consist of BaM and Mg-spinel phase with a molar ratio of 1:2, corresponding to 73.5 wt.% BaM and 26.5 wt.% Mg-spinel (molar mass of Mg-spinel = 200 g, and of BaM = 1111.5 g). These fractions should give a theoretical value of 58.6 emu/g, which is higher than the observed value. The lower observed value could then be attributed to BaM and Mg-spinel phases with saturation magnetizations lower than the theoretical values. The saturation magnetization results are therefore consistent with the picture of phase segregation into BaM and Mg-spinel at sintering temperature of 1100° C. The saturation magnetization of the sample sintered at 1200° C (62.1 emu/g), on the other hand, suggests that the sample consists of about 81.6 wt. % for BaM phase and 18.4 wt. % of $MgFe_2O_4$ phase. This may indicate that a fraction of the Mg ions substitute Fe ions in the hexaferrite phase, which results in a reduction of the mass fraction of the Mg-spinel in the sample.

The coercivity of the samples with $x = 0$ decreased from 1.57 kOe for the sample sintered at 1100° C, down to 0.77 kOe at sintering temperature of 1200° C, and to 0.07 kOe at sintering temperature of 1300° C. The decrease of the coercivity of the sample sintered at 1200° C can be attributed to the significant increase of the particle size as the temperature increases (SEM results). The significant drop in the coercivity of the sample sintered at 1300° C is associated with the crystallization of large hexagonal plates of $Mg_2W$ phase by solid state reaction of the intermediate BaM and $MgFe_2O_4$ phases at high sintering temperatures. The decrease of the squareness ratio ($M_{rs}$) to values below 0.5 (which is characteristic of single-domain, randomly oriented assembly of particles) is consistent with particle growth to the multidomain particle size regime. The multidomain nature of the particles is responsible for the significant decrease in coercivity since in this case, the domain-wall motion is dominant in the magnetization processes.

The saturation magnetization for the sample with $x = 1$ decreased from 64.5 emu/g for the sample sintered at 1100° C, down to 59.84 emu/g at sintering temperature of 1200° C, then increased up to 65.22 emu/g at sintering temperature of 1300° C. The sample sintered at 1100° C is a mixture of BaM phase, and $MgFe_2O_4$ and $CoFe_2O_4$ spinel phases. Again, assuming that the saturation magnetization of the sample sintered at 1100° C results from a simple superposition of the saturation magnetizations of the magnetic phases in the sample, and using the saturation magnetization of 80 emu/g for $CoFe_2O_4$ phase [9], the saturation magnetization for this sample (64.5 emu/g) is consistent with 62 wt. % of BaM phase, 17.5 wt. % of $MgFe_2O_4$ phase, and 20.5 wt. % of $CoFe_2O_4$ phase. In the calculating the weight fractions, the ratio of the molar mass of Co-spinel to that of Mg-spinel (1.173), and equal molar ratios of the two spinel phases were assumed to determine the fractions of the two spinel phases. The relative weight fractions of the BaM and spinel phases (62 wt. % and 38 wt. %, respectively) are similar to those of the sample with $x = 0$ sintered at the same temperature. The saturation magnetization of the sample sintered at 1200° C decreased to 59.84 emu/g), which is consistent with 30 wt.% BaM, 32.2 wt.% Mg-spinel, and 37.8 wt.% Co-spinel. The increase of the weight fraction of the spinel phases (whose weighted average of the saturation magnetization is 55.6 emu/g) cannot be justified on the bases of the starting materials in the sample, and therefore, the reduction of the saturation magnetization could be attributed to the presence of BaM and spinel phases with saturation magnetizations lower than their theoretical values.

The sample sintered at 1300° C is composed of MgCo-W hexaferrite. The increase in saturation magnetization with respect to the sample with $x = 0.0$ can be associated with the increase of the magnetic moment per molecular formula. Since $Mg^{2+}$ is a non-magnetic ion, and the magnetic moment of $Co^{2+}$ ion is $3.7\mu_B$, the increase in the saturation magnetization of this sample suggests that $Co^{2+}$ ions replace $Mg^{2+}$ ions at spin-up sites.

The coercivity decreased from 2.05 kOe for the sample sintered at 1100° C, down to 1.37 kOe at sintering temperature of 1200° C, and to 0.03 kOe at sintering temperature of 1300° C. The decrease of the coercivity of the sample sintered at 1200° C could be associated with the increase of particle size of the BaM phase. The sharp drop in the coercivity of the sample sintered at 1300° C, however, is associated with phase transformation from hard BaM to soft BaW phase with large particle size. The small decrease of the squareness ratio $M_{rs}$ from 0.52 for the sample sintered at 1100° C to 0.45 for the sample sintered at 1200° C is consistent with the small increase of the particle size beyond the critical single domain size as indicated by SEM images. The sharp drop in the squareness ratio down to 0.04 for the sample sintered at 1300° C, however, is due to transformation from hard BaM magnetic phase with almost single-domain particle size to soft, large non-granular masses of BaW phase.

The saturation magnetization for the samples with $x = 2$ revealed similar values at the different sintering temperatures (72.60 emu/g for the sample sintered at 1100° C, 71.23 emu/g at sintering temperature of 1200° C, and 72.56 emu/g at sintering temperature of 1300° C). The sample sintered at 1100° C is a mixture of BaM and $CoFe_2O_4$ as indicated by XRD analysis. The saturation magnetization of 72.6 emu/g is consistent with 74 wt. % BaM and 26 wt. % $CoFe_2O_4$. If we assume that $Co^{2+}$ ions were fully consumed in the spinel phase, the fractions of the BaM and spinel phase would be 70.3 wt.% and 29.7 wt.%, respectively, which correspond to an expected saturation magnetization of 72.96 emu/g. This value is similar to the observed saturation magnetization, which is an indication that this sample consists of BaM and Co-spinel phases with saturation magnetizations close to the theoretical values. On the other hand, the saturation magnetization of 71.23 emu/g for the sample sintered at 1200° C could indicate that Co partially substituted Fe in the hexaferrite lattice. The saturation magnetization of the sample sintered at 1300° C is 72.56 emu/g, which is in good agreement with the previously reported value of 71.18 emu/g for $BaCo_2Fe_{16}O_{27}$ [41], and 74.31 emu/g for $SrCo_2Fe_{16}O_{27}$ [55].

The coercivity decreased from 1.85 kOe for the sample sintered at 1100° C, down to 0.14 kOe at sintering temperature of 1200° C, and down to 0.09 kOe at sintering temperature of 1300° C. These dramatic changes were associated with magnetic softening as a result of particle growth at 1200° C sintering temperature, and the transformation of BaM and Co-spinel phases to large, non-granular masses of W-type hexaferrite phase at 1300° C sintering temperatures. The significant decreases of the squareness ratio $M_{rs}$ from 0.51 to 0.17 and 0.16 is an indication that the samples sintered at 1200° C and 1300° C consist of large multidomain volumes in these samples.

The soft magnetic character, and the low coercivity values of 70 Oe for Mg₂W, 30 Oe for MgCo-W, and 90 Oe for Co₂W for the samples sintered at the highest temperature of 1300° C, are in good agreement with the values reported for Co-substituted Zn₂W ferrites [56]. The increase of the saturation magnetization from 59.44 emu/g for Mg₂W to 65.22 emu/g for CoMg-W, and then to 72.56 emu/g for Co₂W is a result of the substitution of $Mg^{2+}$ ions by $Co^{2+}$ ions at spin-up sites. The saturation trends of the samples (Fig. 12) revealed that the magnetization of sample with $x = 1.0$ is almost saturated at an applied field of 10 kOe, indicating low magnetocrystalline anisotropy relative to the two end compounds ($x = 0.0$ and 2.0).

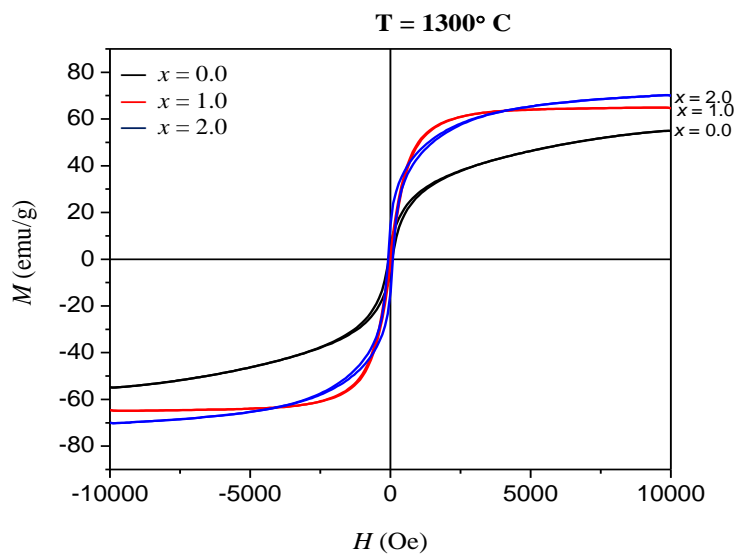

Fig. 12 Hysteresis loops of BaMg₂₋ₓCoₓFe₁₆O₂₇ ($x = 0.0$, 1.0, and 2.0) sintered at 1300° C.

### 3.3.2 **Thermomagnetic measurements**

The thermomagnetic curves of the system $BaMg_{2-x}Co_xFe_{16}O_{27}$ ($x = 0.0$, 1.0, and 2.0) sintered at 1300° C (measured at an applied field of $H = 100$ Oe) are shown in Fig. 13, together with their derivatives with respect to temperature. The derivative curves of all samples exhibited strong negative peaks (dips) at temperatures > 450° C, which were associated with the Curie temperature of the corresponding W-type phase as shown in Table 6. The observed Curie temperature of 452° C for $Mg_2W$ is equal to the previously reported value of (452±3)° C [18], and that of 483° C for $Co_2W$ is in good agreement with the reported values of (490±3)° C [18] and (477±5)° C [46] for this compound, while the Curie temperature of 473° C for MgCo-W lies between the Curie temperatures of the end compounds. The results indicated that the Curie temperature increased with the increase of the Co content, as a consequence of the enhancement of the superexchange interactions resulting from the substitution of $Mg^{2+}$ nonmagnetic ions by $Co^{2+}$ magnetic ions. The occurrence of a single strong dip in the derivative curve suggests that each sample is composed of a single W-type phase.

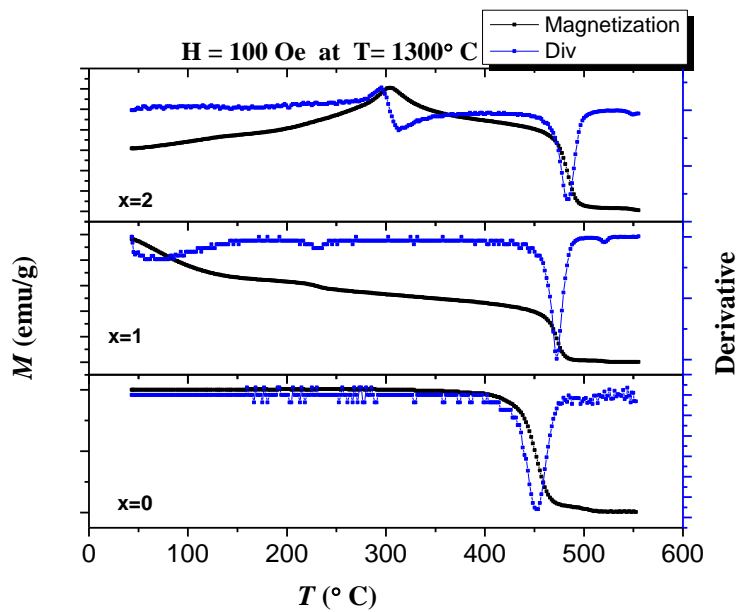

Fig. 13: Thermomagnetic curves of $BaMg_{2-x}Co_xFe_{16}O_{27}$ ($x = 0.0$, 1.0, and 2.0) samples sintered at 1300° C.

**Table 6:** Curie temperature of the different samples $BaMg_{2-x}Co_xFe_{16}O_{27}$ ($x = 0.0$, 1.0, and 2.0).

| $x$ | 0.0 | 1.0 | 2.0 |
|---|---|---|---|
| $Tc$ (°C) | 452 | 473 | 483 |

The derivative curve of the sample with $x = 1.0$ shows two additional weak dips at 232° C and 522° C resulting from inflection points on the magnetization curve. The weak peak at 232° C could be associated with spin reorientation transition, where the magnetic anisotropy changes from easy-cone to easy axis. This temperature is lower than the reported spin reorientation transition temperature for $Co_2W$ [46, 48], which is consistent with the reported increase of the spin reorientation transition temperature with the increase of Co content in Co-substituted $Zn_2W$ ferrite [46, 47]. Further, this result is an indication that both $Mg^{2+}$ and $Co^{2+}$ ions are incorporated into a single W-type phase. The high-temperature dip at 522° C is close to the reported Curie temperature of 527° C for $BaCoZn_{0.5}Mg_{0.5}Fe_{16}O_{27}$ [57], and could be associated with traces of Co-rich magnetic impurities in our sample.

The rise in the thermomagnetic curve of the sample with $x = 2.0$ into a peak at ~ 304° C is associated with the spin reorientation transition from conical to $c$-axis spin orientation. The shape of the derivative curve around this temperature suggests a singular-type behavior of the magnetization, indicating that the origin of the peak is not a smooth function. This behavior is associated with the competition between the rise in the magnetization due to spin reorientation transitions, and the decline due to thermal agitation. The spin reorientation transition in this sample is clearly observed at a higher temperature compared with MgCo-W. Also, a weak negative peak is observed in the derivative curve at ~ 550° C, which can be associated with the enhancement of the superexchange interactions in Co-rich regions of the sample.

## 4. Conclusions

W-type hexaferrites were synthesized by high energy ball milling and sintering at 1300° C. The hexaferrite phase evolved from the reaction of BaM hexaferrite and cubic spinel intermediates which crystallize at lower temperatures. The saturation magnetization as well as the Curie temperature of the W-type hexaferrite increased with the increase of the Co concentration.

Thermomagnetic curves of the W-hexaferrites revealed spin reorientation transition at a temperature which also increased with the increase of the Co content.

## References

[1] R.C. Pullar, Hexagonal ferrites: a review of the synthesis, properties and applications of hexaferrite ceramics, Progress in Materials Science, 57 (2012) 1191-1334.

[2] S.H. Mahmood, Properties and Synthesis of Hexaferrites, in: S.H. Mahmood, I. Abu-Aljarayesh (Eds.) Hexaferrite Permanent Magnetic Materials, Materials Research Forum LLC, Millersville, PA, 2016, pp. 74-110.

[3] S.H. Mahmood, Permanent Magnet Applications, in: S.H. Mahmood, I. Abu-Aljarayesh (Eds.) Hexaferrite Permanent Magnetic Materials, Materials Research Forum LLC, Millersville, PA, 2016, pp. 153-165.

[4] I. Bsoul, S. Mahmood, Structural and magnetic properties of $BaFe_{12-x}Al_xO_{19}$ prepared by milling and calcination, Jordan Journal of Physics, 2 (2009) 171-179.

[5] A. Awadallah, S.H. Mahmood, Y. Maswadeh, I. Bsoul, M. Awawdeh, Q.I. Mohaidat, H. Juwhari, Structural, magnetic, and Mossbauer spectroscopy of Cu substituted M-type hexaferrites, Materials Research Bulletin, 74 (2016) 192-201.

[6] S.H. Mahmood, A. Awadallah, Y. Maswadeh, I. Bsoul, Structural and magnetic properties of Cu-V substituted M-type barium hexaferrites, IOP Conference Series: Materials Science and Engineering, IOP Publishing, 2015, pp. 012008.

[7] A. Awadallah, S.H. Mahmood, Y. Maswadeh, I. Bsoul, A. Aloqaily, Structural and magnetic properties of Vanadium Doped M-Type Barium Hexaferrite ($BaFe_{12-x}V_xO_{19}$), IOP Conference Series: Materials Science and Engineering, IOP Publishing, 2015, pp. 012006.

[8] S.H. Mahmood, I. Bsoul, Hopkinson peak and superparamagnetic effects in $BaFe_{12-x}Ga_xO_{19}$ nanoparticles, EPJ Web of Conferences, 29 (2012) 00039.

[9] J. Smit, H.P.J. Wijn, Ferrites, Wiley, New York, 1959.

[10] S. Chikazumi, Physics of Ferromagnetism 2e, 2nd ed., Oxford University Press Oxford, 2009.

[11] E.S. Alhwaitat, S.H. Mahmood, M. Al-Hussein, O.E. Mohsen, Y. Maswadeh, I. Bsoul, A. Hammoudeh, Effects of synthesis route on the structural and magnetic properties of $Ba_3Zn_2Fe_{24}O_{41}$ ($Zn_2Z$) nanocrystalline hexaferrites, Ceramics International, 44 (2018) 779-787.

[12] S.H. Mahmood, G.H. Dushaq, I. Bsoul, M. Awawdeh, H.K. Juwhari, B.I. Lahlouh, M.A. AlDamen, Magnetic Properties and Hyperfine Interactions in M-Type $BaFe_{12-2x}Mo_xZn_xO_{19}$ Hexaferrites, Journal of Applied Mathematics and Physics, 2 (2014) 77-87.

[13] M. Awawdeh, I. Bsoul, S.H. Mahmood, Magnetic properties and Mössbauer spectroscopy on Ga, Al, and Cr substituted hexaferrites, Journal of Alloys and Compounds, 585 (2014) 465-473.

[14] S. Mahmood, A. Aloqaily, Y. Maswadeh, A. Awadallah, I. Bsoul, H. Juwhari, Structural and magnetic properties of Mo-Zn substituted ($BaFe_{12-4x}Mo_xZn_{3x}O_{19}$) M-type hexaferrites, Material Science Research India, 11 (2014) 09-20.

[15] Y. Maswadeh, S.H. Mahmood, A. Awadallah, A.N. Aloqaily, Synthesis and structural characterization of nonstoichiometric barium hexaferrite materials with Fe: Ba ratio of 11.5–16.16, IOP Conference Series: Materials Science and Engineering, IOP Publishing, 2015, pp. 012019.

[16] S.H. Mahmood, A.N. Aloqaily, Y. Maswadeh, A. Awadallah, I. Bsoul, M. Awawdeh, H.K. Juwhari, Effects of heat treatment on the phase evolution, structural, and magnetic properties of Mo-Zn doped M-type hexaferrites, Solid State Phenomena, 232 (2015) 65-92.

[17] A. Alsmadi, I. Bsoul, S. Mahmood, G. Alnawashi, K. Prokeš, K. Siemensmeyer, B. Klemke, H. Nakotte, Magnetic study of M-type doped barium hexaferrite nanocrystalline particles, Journal of Applied Physics, 114 (2013) 243910.

[18] A. Collomb, P. Wolfers, X. Obradors, Neutron diffraction studies of some hexagonal ferrites: $BaFe_{12}O_{19}$, $BaMg_2$-W and $BaCo_2$-W, Journal of magnetism and magnetic materials, 62 (1986) 57-67.

[19] M. Ahmad, R. Grössinger, M. Kriegisch, F. Kubel, M. Rana, Magnetic and microwave attenuation behavior of Al-substituted $Co_2$W hexaferrites synthesized by sol-gel autocombustion process, Current Applied Physics, 12 (2012) 1413-1420.

[20] M.J. Iqbal, R.A. Khan, S. Mizukami, T. Miyazaki, Mössbauer and magnetic study of Mn, Zr and Cd substituted W-type hexaferrites prepared by co-precipitation, Materials Research Bulletin, 46 (2011) 1980-1986.

[21] F. Leccabue, O.A. Muzio, M.S.E. Kany, G. Calestani, G. Albanese, Magnetic properties and phase formation of $SrMn_2Fe_{16}O_{27}$ ($SrMn_2$-W) hexaferrite prepared by the coprecipitation method, Journal of magnetism and magnetic materials, 68 (1987) 201-212.

[22] M.J. Iqbal, R.A. Khan, Enhancement of electrical and dielectric properties of Cr doped $BaZn_2$W-type hexaferrite for potential applications in high frequency devices, Journal of Alloys and Compounds, 478 (2009) 847-852.

[23] Y. Yang, B. Zhang, W. Xu, Y. Shi, N. Zhou, H. Lu, Microwave absorption studies of W-hexaferrite prepared by co-precipitation/mechanical milling, Journal of magnetism and magnetic materials, 265 (2003) 119-122.

[24] M. Ahmad, F. Aen, M. Islam, S.B. Niazi, M. Rana, Structural, physical, magnetic and electrical properties of La-substituted W-type hexagonal ferrites, Ceramics International, 37 (2011) 3691-3696.

[25] X. Qin, Y. Cheng, K. Zhou, S. Huang, X. Hui, Microwave Absorbing Properties of W-Type Hexaferrite Ba $(MnZn)_xCo_{2(1-x)}Fe_{16}O_{27}$, Journal of Materials Science and Chemical Engineering, 1 (2013) 8.

[26] Z. Zi, J. Dai, Q. Liu, H. Liu, X. Zhu, Y. Sun, Magnetic and microwave absorption properties of W-type $Ba(Zn_xCo_{1-x})_2Fe_{16}O_{27}$ hexaferrite platelets, Journal of Applied Physics, 109 (2011) 07E536.

[27] M.J. Iqbal, R.A. Khan, S. Mizukami, T. Miyazaki, Tailoring of structural, electrical and magnetic properties of $BaCo_2$W-type hexaferrites by doping with Zr–Mn binary mixtures for useful applications, Journal of Magnetism and Magnetic Materials, 323 (2011) 2137-2144.

[28] G. Shen, Z. Xu, Y. Li, Absorbing properties and structural design of microwave absorbers based on W-type La-doped ferrite and carbon fiber composites, Journal of magnetism and magnetic materials, 301 (2006) 325-330.

[29] J. Xu, H. Zou, H. Li, G. Li, S. Gan, G. Hong, Influence of $Nd^{3+}$ substitution on the microstructure and electromagnetic properties of barium W-type hexaferrite, Journal of Alloys and Compounds, 490 (2010) 552-556.

[30] L. Deng, L. Ding, K. Zhou, S. Huang, Z. Hu, B. Yang, Electromagnetic properties and microwave absorption of W-type hexagonal ferrites doped with $La^{3+}$, Journal of Magnetism and Magnetic Materials, 323 (2011) 1895-1898.

[31] M. Qiao, C. Zhang, H. Jia, Synthesis and absorbing mechanism of two-layer microwave absorbers containing flocs-like nano-$BaZn_{1.5}Co_{0.5}Fe_{16}O_{27}$ and carbonyl iron, Materials Chemistry and Physics, 135 (2012) 604-609.

[32] Y. Wu, Y. Huang, L. Niu, Y. Zhang, Y. Li, X. Wang, $Pr^{3+}$-substituted W-type barium ferrite: Preparation and electromagnetic properties, Journal of Magnetism and Magnetic Materials, 324 (2012) 616-621.

[33] M. Ahmad, R. Grössinger, I. Ali, I. Ahmad, M. Rana, Synthesis and characterization of Al-substituted W-type hexagonal ferrites for high frequency applications, Journal of Alloys and Compounds, 577 (2013) 382-388.

[34] M. Ahmad, I. Ali, R. Grössinger, M. Kriegisch, F. Kubel, M. Rana, Effects of divalent ions substitution on the microstructure, magnetic and electromagnetic parameters of $Co_2W$ hexagonal ferrites synthesized by sol–gel method, Journal of Alloys and Compounds, 579 (2013) 57-64.

[35] F. Aen, M. Ahmad, M. Rana, The role of Ga substitution on magnetic and electromagnetic properties of nano-sized W-type hexagonal ferrites, Current Applied Physics, 13 (2013) 41-46.

[36] Y. Wu, C. Ong, G. Lin, Z. Li, Improved microwave magnetic and attenuation properties due to the dopant $V_2O_5$ in W-type barium ferrites, Journal of Physics D: Applied Physics, 39 (2006) 2915.

[37] L. Wang, J. Song, Q. Zhang, X. Huang, N. Xu, The microwave magnetic performance of $Sm^{3+}$ doped $BaCo_2Fe_{16}O_{27}$, Journal of Alloys and Compounds, 481 (2009) 863-866.

[38] M.J. Iqbal, R.A. Khan, S. Mizukami, T. Miyazaki, Mössbauer, magnetic and microwave absorption characteristics of substituted W-type hexaferrites nanoparticles, Ceramics International, 38 (2012) 4097-4103.

[39] I. Khan, I. Sadiq, M.N. Ashiq, Role of Ce–Mn substitution on structural, electrical and magnetic properties of W-type strontium hexaferrites, Journal of Alloys and Compounds, 509 (2011) 8042-8046.

[40] S. Ram, Crystallization of acicular platelet particles of W-type hexagonal strontium ferrite for magnetic recording applications, Journal of Materials Science, 25 (1990) 2465-2470.

[41] F. Guo, X. Wu, G. Ji, J. Xu, L. Zou, S. Gan, Synthesis and Properties Investigation of Non-equivalent Substituted W-Type Hexaferrite, Journal of Superconductivity and Novel Magnetism, 27 (2014) 411-420.

[42] Z. Su, Y. Chen, B. Hu, A.S. Sokolov, S. Bennett, L. Burns, X. Xing, V.G. Harris, Crystallographically textured self-biased W-type hexaferrites for X-band microwave applications, Journal of Applied Physics, 113 (2013) 17B305.

[43] Y. Feng, T. Qiu, C. Shen, Absorbing properties and structural design of microwave absorbers based on carbonyl iron and barium ferrite, Journal of Magnetism and Magnetic Materials, 318 (2007) 8-13.

[44] A. Pasko, F. Mazaleyrat, M. Lobue, V. Loyau, V. Basso, M. Küpferling, C. Sasso, L. Bessais, Magnetic and structural characterization of nanosized $BaCo_xZn_{2-x}Fe_{16}O_{27}$ hexaferrite in

the vicinity of spin reorientation transition, Journal of Physics: conference series, IOP Publishing, 2011, pp. 012045.

[45] S. Rinaldi, F. Licci, A. Paoluzi, G. Turilli, Competing anisotropies and first-order magnetization processes in (Zn,Co) W-type hexaferrite, Journal of Applied Physics, 60 (1986) 3680-3684.

[46] G. Albanese, E. Calabrese, A. Deriu, F. Licci, Mössbauer investigation of W-type hexaferrite of composition $BaZn_{2-x}Co_xFe_{16}O_{27}$, Hyperfine Interactions, 28 (1986) 487-489.

[47] A. Paoluzi, F. Licci, O. Moze, G. Turilli, A. Deriu, G. Albanese, E. Calabrese, Magnetic, Mössbauer, and neutron diffraction investigations of W-type hexaferrite $BaZn_{2-x}Co_xFe_{16}O_{27}$ single crystals, Journal of applied physics, 63 (1988) 5074-5080.

[48] D. Samaras, A. Collomb, S. Hadjivasiliou, C. Achilleos, J. Tsoukalas, J. Pannetier, J. Rodriguez, The rotation of the magnetization in the $BaCo_2Fe_{16}O_{27}$ W-type hexagonal ferrite, Journal of magnetism and magnetic materials, 79 (1989) 193-201.

[49] R.A. Khan, S. Mir, A.M. Khan, B. Ismail, A.R. Khan, Doping magnesium ion to tune electrical and dielectric properties of $BaCo_2$ hexaferrites, Ceramics International, 40 (2014) 11205-11211.

[50] R.t. Shannon, Revised effective ionic radii and systematic studies of interatomic distances in halides and chalcogenides, Acta Crystallographica Section A: Crystal Physics, Diffraction, Theoretical and General Crystallography, 32 (1976) 751-767.

[51] C. Suryanarayana, M.G. Norton, X-Ray Diffraction: A Practical Approach, Springer Science & Business Media1998.

[52] L. Rezlescu, E. Rezlescu, P. Popa, N. Rezlescu, Fine barium hexaferrite powder prepared by the crystallisation of glass, Journal of Magnetism and Magnetic Materials, 193 (1999) 288-290.

[53] B.D. Cullity, C.D. Graham, Introduction to magnetic materials, 2nd ed., John Wiley & Sons, Hoboken, NJ, 2011.

[54] M. Ahmad, R. Grössinger, M. Kriegisch, F. Kubel, M. Rana, Characterization of Sr-substituted W-type hexagonal ferrites synthesized by sol–gel autocombustion method, Journal of Magnetism and Magnetic Materials, 332 (2013) 137-145.

[55] C. Stergiou, G. Litsardakis, Preparation and magnetic characterization of $Co_2$-W strontium hexaferrites doped with Ni and La, Journal of Magnetism and Magnetic Materials, 323 (2011) 2362-2368.

[56] D. Hemeda, A. Al-Sharif, O. Hemeda, Effect of Co substitution on the structural and magnetic properties of Zn–W hexaferrite, Journal of magnetism and magnetic materials, 315 (2007) L1-L7.

[57] M. Ahmed, N. Okasha, M. Oaf, R. Kershi, The role of Mg substitution on the microstructure and magnetic properties of Ba Co Zn W-type hexagonal ferrites, Journal of magnetism and magnetic materials, 314 (2007) 128-134.